\documentclass[letterpaper, 10 pt, conference]{ieeeconf}  

\IEEEoverridecommandlockouts                              

\usepackage[]{graphicx}    
\usepackage[colorlinks=true, urlcolor=blue, linkcolor=black, citecolor=black]{hyperref} 
\usepackage{newtxtext}
\usepackage{amsmath}
\usepackage{booktabs}
\usepackage{amssymb}
\usepackage{subcaption}
\usepackage{placeins}
\usepackage{multirow}

\title{\LARGE \bf
Augmented Model Predictive Control: A Balance between Satellite Agility and Computation Complexity
}

\author{Yiming Wang$^{1,2}$ and Mihindukulasooriya Sheral Crescent Tissera$^{2}$ and Haihong Yu$^{2}$ \\and Kai Jie Ethan Foo$^{2}$ and Sean Yeo Keyuan$^{2}$ and Ankit Srivastava$^{2}$ and Hao An$^{1}$
\thanks{$^{1}$ Department of Control Science and Engineering, Harbin Institute of Technology, Harbin, China.
        {\tt\small 22b904002@stu.hit.edu.cn}}%
\thanks{$^{2}$ Department of Electrical and Computer Engineering, National University of Singapore, Singapore.
        {\tt\small tsheral@nus.edu.sg}}%
}

\begin{document}

\maketitle
\thispagestyle{empty}
\pagestyle{empty}

\begin{abstract}
Agile earth observation satellites employ multiple actuators to enable flexible and responsive imaging capabilities.
While significant advancements in actuator technology have enhanced satellites' torque and momentum, relatively little attention has been given to control strategies specifically tailored to improve satellite agility.
This paper provides a comparative analysis of different Model Predictive Control (MPC) formulations and introduces an augmented-MPC method that effectively balances agility requirements with hardware implementation constraints.
The proposed method achieves the high-performance characteristics of nonlinear MPC while preserving the computational simplicity of linear MPC.
Numerical simulations and physical experiments are conducted to validate the effectiveness and feasibility of the proposed approach.
\end{abstract}

\section{INTRODUCTION}
Earth Observation Satellites (EOSs) are spacecraft engineered to remotely monitor the Earth's natural environment and human activities from orbit.
Equipped with diverse sensors—including visible light cameras, multispectral imagers, and thermal infrared detectors—these satellites collect critical data about the Earth's surface and the near-Earth atmosphere \cite{intro2}.
They serve as indispensable tools in various domains such as resource exploration, urban planning, environmental surveillance, and disaster assessment.

In recent years, small satellites weighing less than 50 kg and with dimensions smaller than a laser printer have gained widespread popularity.
This trend is driven by advancements in commercial off-the-shelf technologies and the adoption of agile development approaches.
These satellites have opened up space as the next frontier for civilization and are part of what the space industry has called "NewSpace", referring to the increased accessibility for commercialization in the private space sector and the recent exponential progress in the nano and micro satellite industry \cite{intro1}.

To broaden their observational capabilities, Agile EOSs (AEOSs) are outfitted with multiple actuators—such as reaction wheels and control moment gyroscopes—that enable full three-degree-of-freedom attitude control \cite{actuator1}.
This agility allows AEOSs to perform off-nadir observations by adjusting their orientation, rather than relying solely on direct overhead passes \cite{scheduling1}.
As a result, AEOSs greatly improve the practicality of satellite scheduling problems, where observation tasks must be efficiently assigned within constrained time windows to maximize overall mission value \cite{scheduling2}.

Agility, in the context of satellite operations, refers to the ability to rapidly reorient in space.
A satellite that can swiftly transition between tracking targets and maintain stable pointing performance is inherently more agile.
Consequently, higher agility translates into enhanced reliability and flexibility in executing complex observation schedules.
In this context, improving satellite agility is essential to meet the increasing demand for rapid retargeting and time-sensitive space missions.

Recent advancements have focused heavily on enhancing actuator performance—particularly by increasing the maximum torque and angular momentum output—which is fundamental to improving satellite agility.
To enhance such agility, NewSpace companies have launched high-torque-producing actuators such as miniaturized control moment gyroscopes.
However, due to strict size, weight, and power constraints associated with small satellites, enhancing agility through hardware alone is not optimal.
Thus, a critical yet frequently overlooked challenge is how to maximize the agility of satellites using low-power, cost-effective actuators.
This leads to a key question: can advanced control strategies boost satellite agility without relying solely on hardware upgrades?

Model Predictive Control (MPC) has garnered significant attention in both academic and industrial communities for its ability to balance control precision and energy efficiency, particularly under actuator saturation constraints.
Among its variants, Linear MPC (LMPC) is widely used in satellite systems due to its relatively low computational demands.
For example, Zhao et al. highlighted the high computational requirements of MPC and proposed an advanced attitude controller based on LMPC \cite{LMPC1}.
similarly, Caverly et al. employed LMPC to retain low computational burden \cite{LMPC2}.
In contrast, Nonlinear MPC (NMPC) captures the system's nonlinear dynamics more accurately and offers superior performance by optimizing a more realistic cost function \cite{add20250909}.
Zhou et al. integrated an online model identification algorithm with NMPC and used Sequential Quadratic Programming (SQP) to solve the resulting Optimal Control Problem (OCP) \cite{nmpc1}.
Song et al. demonstrated that once an optimization objective is defined, NMPC with SQP can outperform reinforcement learning in control tasks \cite{RLvsOC}.
Kamel et al. also reported that NMPC provides greater robustness and dynamic performance than LMPC in comparative experiments \cite{LMPCvsNMPC}.
Nevertheless, these benefits come with a significantly higher computational cost, as the complexity of solving NMPC increases rapidly with the prediction horizon. This makes NMPC highly demanding in terms of onboard processing power \cite{ECC1,ECC2}.

As reviewed in the literature, LMPC is generally preferred over NMPC for satellite attitude control due to its lower computational requirements.
The high computational load imposed by NMPC is particularly problematic for small satellites, leading to increased system complexity and cost.
Moreover, NMPC often fails to achieve real-time performance because of its iterative optimization process, which can introduce unacceptable delays in time-sensitive aerospace missions.
On the other hand, LMPC relies on simplified models, resulting in suboptimal solutions compared to NMPC and thus offering reduced agility.
This presents a compelling challenge: how can one retain the superior performance of NMPC while preserving the computational efficiency of LMPC?
To the best of the authors' knowledge, no effective solution has been proposed. This paper addresses this gap by introducing an augmented-LMPC approach.
By embedding an integrator into the baseline LMPC structure, the proposed method improves tracking agility without necessitating changes to the actuator hardware.

The remainder of this paper is structured as follows:
Section \ref{cpt2} presents the satellite kinematic and dynamic models, along with the baseline LMPC framework.
Section \ref{cpt3} introduces the proposed augmented-LMPC method and formulates several OCPs for different MPC methods.
Section \ref{cpt4} provides comprehensive comparative experiments to evaluate the performance of different MPC methods.
Finally, Section \ref{cpt5} concludes the paper with a summary of findings and potential future directions.

\section{Preliminaries}\label{cpt2}
In this section, the nonlinear kinematic and dynamic models of the satellite are formulated.
Based on these complete models, system linearization is performed.

\subsection{Nonlinear Model}
To avoid the singularity inherent in Euler angles, this paper employs quaternions to represent the satellite's attitude.  
The Earth-Centered Inertial (ECI) frame is selected as the inertial reference frame.
Based on the satellite's orbital position, the quaternion representing the rotation from the ECI frame to the Local-Vertical/Local-Horizontal (LVLH) frame is first computed.
This quaternion is then further rotated to align with the target frame, according to the target's relative latitude and longitude, yielding the target quaternion $\boldsymbol{q}_t \in \mathbb{R}^4$.

The attitude kinematics and dynamics of the satellite are given as follows \cite{book1}:
\begin{equation}\label{eq1}
\dot{\boldsymbol{q}}=\frac{1}{2}
\left[\begin{array}{rrrr}
q_0&-q_1&-q_2&-q_3\\
q_1&q_0&-q_3&q_2\\
q_2&q_3&q_0&-q_1\\
q_3&-q_2&q_1&q_0
\end{array}\right]
\left[\begin{array}{c}
0\\
\boldsymbol{\omega}
\end{array}\right]
\end{equation}
\begin{equation}\label{eq2}
\dot{\boldsymbol{\omega}}=\boldsymbol{J}^{-1} \left(\boldsymbol{T}-\boldsymbol{\omega}\times \left(\boldsymbol{J}\boldsymbol{\omega}+\boldsymbol{h}_{rw}\right) \right)
\end{equation}
where $\boldsymbol{q}=\left[q_0, q_1, q_2, q_3\right]^T$ denotes the satellite's attitude quaternion,  
$\boldsymbol{\omega}=\left[P, Q, R\right]^T$ is the angular rate,  
$\boldsymbol{T}\in\mathbb{R}^{3}$ represents the control torque generated by reaction wheels,  
and $\boldsymbol{h}_{rw}\in\mathbb{R}^{3}$ is the reaction wheel momentum, which satisfies the following relation:
\begin{equation}\label{eq3}
\dot{\boldsymbol{h}}_{rw}=-\boldsymbol{T}
\end{equation}
Here, $\boldsymbol{J}\in\mathbb{R}^{3\times 3}$ is the satellite's inertia matrix, with the values listed in Table \ref{table1}.
It should be noted that the disturbances are not included in equation (\ref{eq2}) for controller derivation, but can be included in the system dynamics with a reaction wheel momentum dumping controller running in parallel.

\begin{table}[!t]
\renewcommand{\arraystretch}{1.3}
\caption{\textbf{Satellite Moment of Inertia}}
\label{table1}
\centering
\begin{tabular}{|c|c|c|}
\hline
\bfseries Parameter & \bfseries Value  & \bfseries Unit\\
\hline\hline
$J_{xx}$ & 0.2912908 & $\text{kg}\cdot \text{m}^2$\\
$J_{yy}$ & 0.2837495 & $\text{kg}\cdot \text{m}^2$\\
$J_{zz}$ & 0.3940411 & $\text{kg}\cdot \text{m}^2$\\
$J_{xy}$ & -0.0024154 & $\text{kg}\cdot \text{m}^2$\\
$J_{xz}$ & 0.0011626 & $\text{kg}\cdot \text{m}^2$\\
$J_{yz}$ & 0.0009412 & $\text{kg}\cdot \text{m}^2$\\
\hline
\end{tabular}
\end{table}

The quaternion error $\boldsymbol{q}_e\in\mathbb{R}^{4}$ is defined as
\begin{equation}\label{eq4}
\boldsymbol{q}_e=\boldsymbol{q}\otimes\boldsymbol{q}_t^{-1}
\end{equation}
where the operator ``$\otimes$'' denotes quaternion multiplication.

\subsection{Linearization}
The error quaternion $\boldsymbol{q}_e$ can be separated into two components, and the kinematic equation (\ref{eq1}) is rewritten as
\begin{equation}\label{eq5}
\dot{\boldsymbol{q}}_e=
\left[\begin{array}{c}
\dot{\eta}\\
\dot{\boldsymbol{\xi}}
\end{array}\right]
=\frac{1}{2}
\left[\begin{array}{c}
-\boldsymbol{\xi}^T\\
\eta \boldsymbol{I}_{3\times3}+\left[\boldsymbol{\xi}\times\right]
\end{array}\right]\boldsymbol{\omega}
\end{equation}
where $\eta$ denotes the scalar part, $\boldsymbol{\xi}=[\xi_1, \xi_2, \xi_3]^T$ is the vector part,  
$\boldsymbol{I}_{3\times3}$ is the three-dimensional identity matrix, and $\left[\boldsymbol{\xi}\times\right]\in\mathbb{R}^{3\times 3}$ is the skew-symmetric matrix corresponding to the cross-product operation:
\begin{equation}\label{eq6}
\left[\boldsymbol{\xi}\times\right]=
\left[\begin{array}{ccc}
0 & -\xi_3 & \xi_2 \\
\xi_3 & 0 & -\xi_1 \\
-\xi_2 & \xi_1 & 0
\end{array}\right]
\end{equation}

Applying a Taylor expansion around the equilibrium point $\boldsymbol{\xi}=\boldsymbol{0}$ and neglecting higher-order terms yields the linearized form \cite{ULMPC}:
\begin{equation}\label{eq7}
\dot{\boldsymbol{\xi}}=\tfrac{1}{2}\boldsymbol{I}_{3\times3}\boldsymbol{\omega}
\end{equation}

Next, the dynamic equation (\ref{eq2}) is reformulated as
\begin{equation}\label{eq8}
\dot{\boldsymbol{\omega}}=\boldsymbol{J}^{-1}\boldsymbol{T}_c
\end{equation}
where $\boldsymbol{T}_c\in\mathbb{R}^{3}$ denotes the feedforward control torque.  
The reaction wheel torque command is then expressed as
\begin{equation}\label{eq9}
\boldsymbol{T}=\boldsymbol{T}_c+\boldsymbol{\omega}\times \left(\boldsymbol{J}\boldsymbol{\omega}+\boldsymbol{h}_{rw}\right)
\end{equation}

Combining the linearized kinematic and dynamic models (\ref{eq7}--\ref{eq8}), the satellite attitude error dynamics in state-space form can be written as
\begin{equation}\label{eq10}
\left\{
\begin{array}{l}
\dot{\boldsymbol{x}}=\boldsymbol{A}_1\boldsymbol{x}+\boldsymbol{B}_1\boldsymbol{T}_c\\
\boldsymbol{y}=\boldsymbol{C}_1\boldsymbol{x}
\end{array}
\right.
\end{equation}
where $\boldsymbol{x}=[\boldsymbol{\xi}; \boldsymbol{\omega}]$,  
$\boldsymbol{y}=\boldsymbol{\xi}$, and the system matrices are given by
\begin{equation}\label{eq11}
\begin{aligned}
&\boldsymbol{A}_1=
\left[\begin{array}{cc}
\boldsymbol{0}_{3\times3} & \tfrac{1}{2}\boldsymbol{I}_{3\times3}\\
\boldsymbol{0}_{3\times3} & \boldsymbol{0}_{3\times3}
\end{array}\right],\quad
\boldsymbol{B}_1=
\left[\begin{array}{c}
\boldsymbol{0}_{3\times3}\\
\boldsymbol{J}^{-1}
\end{array}\right],\\
&\boldsymbol{C}_1=
\left[\boldsymbol{I}_{3\times3},\; \boldsymbol{0}_{3\times3}\right]
\end{aligned}
\end{equation}

For a sampling time $t_s$, the discrete-time state-space model corresponding to (\ref{eq10}) can be expressed as \cite{book2}
\begin{equation}\label{eq12}
\left\{
\begin{array}{l}
\boldsymbol{x}(k+1)=\boldsymbol{A}_2\boldsymbol{x}(k)+\boldsymbol{B}_2\boldsymbol{T}_c(k)\\
\boldsymbol{y}(k)=\boldsymbol{C}_2\boldsymbol{x}(k)
\end{array}
\right.
\end{equation}
with matrices defined as
\begin{equation}\label{eq13}
\boldsymbol{A}_2=e^{\boldsymbol{A}_1t_s},\quad
\boldsymbol{B}_2=\left(\int_0^{t_s} e^{\boldsymbol{A}_1t}\,\textrm{d}t\right)\boldsymbol{B}_1,\quad
\boldsymbol{C}_2=\boldsymbol{C}_1
\end{equation}

\section{Methodology}\label{cpt3}
To systematically introduce different MPC strategies for satellite attitude control, we begin with the augmented model and then present several MPC formulations.

\subsection{Augmented Model}
Based on the discrete-time state-space equations (\ref{eq12}), the augmented model is formulated as \cite{book2}
\begin{equation}\label{eq14}
\left\{
\begin{array}{l}
\boldsymbol{x}_a(k+1)=\boldsymbol{A}_a\boldsymbol{x}_a(k)+\boldsymbol{B}_a\boldsymbol{\Delta T}_c(k)\\
\boldsymbol{y}_a(k)=\boldsymbol{C}_a\boldsymbol{x}_a(k)
\end{array}
\right.
\end{equation}
where $\boldsymbol{x}_a(k)=[\boldsymbol{\Delta x}(k); \boldsymbol{y}_a(k)]$ and $\boldsymbol{y}_a(k)=\boldsymbol{y}(k)$.  
The operator $\boldsymbol{\Delta}$ denotes the difference operator, with $\boldsymbol{\Delta x}(k)=\boldsymbol{x}(k)-\boldsymbol{x}(k-1)$ and $\boldsymbol{\Delta T}_c(k)=\boldsymbol{T}_c(k)-\boldsymbol{T}_c(k-1)$ representing the increment of state variables and control torques, respectively.  
The corresponding matrices are expressed as
\begin{equation}\label{eq15}
\begin{aligned}
&\boldsymbol{A}_a=
\left[\begin{array}{cc}
\boldsymbol{A}_2&\boldsymbol{0}_{6\times3}\\
\boldsymbol{C}_2\boldsymbol{A}_2&\boldsymbol{I}_{3\times3}
\end{array}\right],\quad
\boldsymbol{B}_a=
\left[\begin{array}{c}
\boldsymbol{B}_2\\
\boldsymbol{C}_2\boldsymbol{B}_2
\end{array}\right],\\
&\boldsymbol{C}_a=
\left[\boldsymbol{0}_{3\times6}, \; \boldsymbol{I}_{3\times3}\right]
\end{aligned}
\end{equation}

\subsection{Augmented-Constrained LMPC (A-CLMPC)}
From the augmented model (\ref{eq14}), the predicted output over the horizon is obtained as
\begin{equation}\label{eq16}
\boldsymbol{\mathcal{Y}}_a(k)=\boldsymbol{\Phi}_a\boldsymbol{x}_a(k)+\boldsymbol{R}_a\boldsymbol{\Delta\mathcal{T}}_c(k)
\end{equation}
where $\boldsymbol{\mathcal{Y}}_a(k)=[\boldsymbol{y}_a(k+1); \boldsymbol{y}_a(k+2); \cdots; \boldsymbol{y}_a(k+N_p)]$, with $N_p$ denoting the prediction horizon.  
The control sequence is $\boldsymbol{\Delta\mathcal{T}}_c(k)=[\boldsymbol{\Delta T}_c(k); \boldsymbol{\Delta T}_c(k+1); \cdots; \boldsymbol{\Delta T}_c(k+N_c-1)]$, where $N_c$ is the control horizon and $N_c \leqslant N_p$.  
The matrices are expanded as
\begin{multline}\label{eq17}
\boldsymbol{\Phi}_a=
\left[\begin{array}{c}
\boldsymbol{C}_a\boldsymbol{A}_a\\
\boldsymbol{C}_a\boldsymbol{A}_a^2\\
\vdots\\
\boldsymbol{C}_a\boldsymbol{A}_a^{N_p}
\end{array}\right],\\
\boldsymbol{R}_a=
\left[\begin{array}{cc}
\boldsymbol{C}_a\boldsymbol{B}_a&\boldsymbol{0}_{3\times3}\\
\boldsymbol{C}_a\boldsymbol{A}_a\boldsymbol{B}_a&\boldsymbol{C}_a\boldsymbol{B}_a\\
\vdots&\vdots\\
\boldsymbol{C}_a\boldsymbol{A}_a^{N_p-1}\boldsymbol{B}_a&\boldsymbol{C}_a\boldsymbol{A}_a^{N_p-2}\boldsymbol{B}_a
\end{array}\right.\\
\left.\begin{array}{cc}
\cdots&\boldsymbol{0}_{3\times3}\\
\cdots&\boldsymbol{0}_{3\times3}\\
\ddots&\vdots\\
\cdots&\boldsymbol{C}_a\boldsymbol{A}_a^{N_p-N_c}\boldsymbol{B}_a
\end{array}\right]
\end{multline}

The associated cost function is defined as
\begin{multline}\label{eq18}
J(k)=\boldsymbol{\mathcal{Y}}_a^T(k)\boldsymbol{\mathcal{Y}}_a(k) \\
\hfill + W_{A-CLMPC}\boldsymbol{\Delta\mathcal{T}}_c^T(k)\boldsymbol{\Delta\mathcal{T}}_c(k)
\end{multline}
where $W_{A-CLMPC}$ is a weight to balance tracking accuracy and control energy.  
Thus, the satellite attitude tracking problem can be formulated as the following OCP:
\begin{equation}\label{eq19}
\begin{aligned}
\operatorname*{argmin}_{\boldsymbol{\Delta\mathcal{T}}_c(k)} \quad & J(k) \\
\text{s.t.} \quad & \boldsymbol{T}_{min} \leqslant \boldsymbol{T}_c(k+i) \leqslant \boldsymbol{T}_{max},\\
&i=0, 1, \cdots, N_c-1\\
                 & \boldsymbol{h}_{min} \leqslant \boldsymbol{h}_{rw}(k)-t_s\sum_{j=0}^{i}\boldsymbol{T}_c(k+j) \leqslant \boldsymbol{h}_{max},\\
&i=0, 1, \cdots, N_c-1
\end{aligned}
\end{equation}
Here, $\boldsymbol{T}_{min}$ and $\boldsymbol{T}_{max}$ denote torque limits of the reaction wheels, while $\boldsymbol{h}_{min}$ and $\boldsymbol{h}_{max}$ represent the corresponding momentum bound.  
In this paper, the quadratic programming is solved using qpOASES \cite{qpOASES}.  
Following the receding horizon strategy of MPC, only the first set of computed torque is applied at each step.

For further performance comparison, several alternative MPC formulations and their corresponding OCPs are summarized below.

\subsection{Unconstrained LMPC (ULMPC)}
In our previous work \cite{ULMPC}, to minimize algorithmic complexity, torque and momentum constraints of the reaction wheel were omitted.
From the discrete-time model (\ref{eq12}), the predicted output is given by
\begin{equation}\label{eq20}
\boldsymbol{\mathcal{Y}}(k)=\boldsymbol{\Phi}\boldsymbol{x}(k)+\boldsymbol{R}\boldsymbol{\mathcal{T}}_c(k)
\end{equation}
where $\boldsymbol{\mathcal{Y}}(k)=[\boldsymbol{y}(k+1); \boldsymbol{y}(k+2); \cdots; \boldsymbol{y}(k+N_p)]$ and  
$\boldsymbol{\mathcal{T}}_c(k)=[\boldsymbol{T}_c(k); \boldsymbol{T}_c(k+1); \cdots; \boldsymbol{T}_c(k+N_c-1)]$.  
The matrices are defined as
\begin{multline}\label{eq21}
\boldsymbol{\Phi}=
\left[\begin{array}{c}
\boldsymbol{C}_2\boldsymbol{A}_2\\
\boldsymbol{C}_2\boldsymbol{A}_2^2\\
\vdots\\
\boldsymbol{C}_2\boldsymbol{A}_2^{N_p}
\end{array}\right],\\
\boldsymbol{R}=
\left[\begin{array}{cc}
\boldsymbol{C}_2\boldsymbol{B}_2&\boldsymbol{0}_{3\times3}\\
\boldsymbol{C}_2\boldsymbol{A}_2\boldsymbol{B}_2&\boldsymbol{C}_2\boldsymbol{B}_2\\
\vdots&\vdots\\
\boldsymbol{C}_2\boldsymbol{A}_2^{N_p-1}\boldsymbol{B}_2&\boldsymbol{C}_2\boldsymbol{A}_2^{N_p-2}\boldsymbol{B}_2
\end{array}\right.\\
\left.\begin{array}{cc}
\cdots&\boldsymbol{0}_{3\times3}\\
\cdots&\boldsymbol{0}_{3\times3}\\
\ddots&\vdots\\
\cdots&\boldsymbol{C}_2\boldsymbol{A}_2^{N_p-N_c}\boldsymbol{B}_2
\end{array}\right]
\end{multline}

The corresponding OCP is
\begin{equation}\label{eq22}
\operatorname*{argmin}_{\boldsymbol{\mathcal{T}}_c(k)} \quad  \boldsymbol{\mathcal{Y}}^T(k)\boldsymbol{\mathcal{Y}}(k)+W_{ULMPC}\boldsymbol{\mathcal{T}}_c^T(k)\boldsymbol{\mathcal{T}}_c(k)
\end{equation}

\begin{figure*}[t]
\vspace{3mm}
\centering
\includegraphics[width=\textwidth]{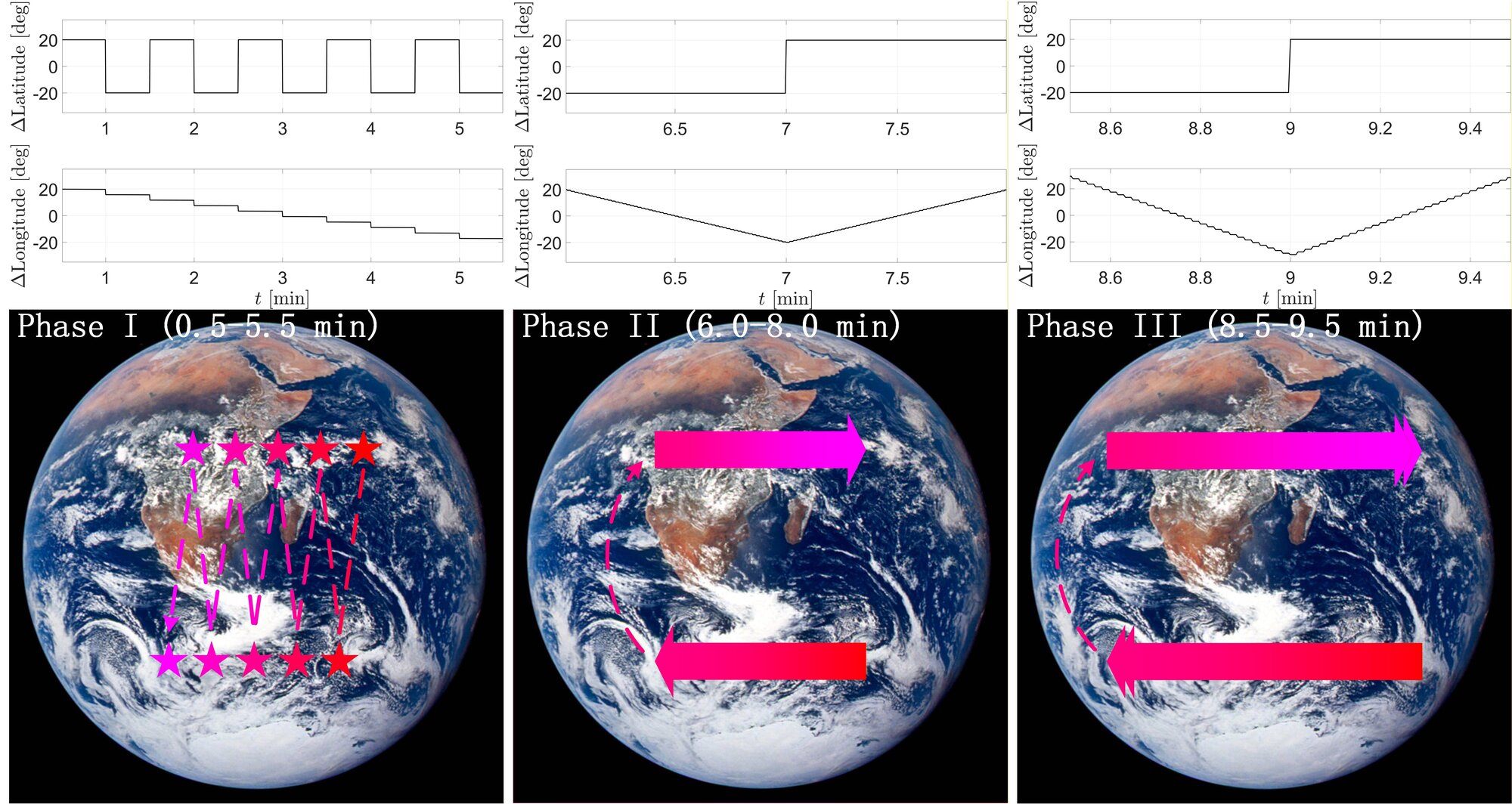}
\caption{Target latitude and longitude offsets in three phases.}
\label{fig1}
\end{figure*}

\subsection{CLMPC}
Extending ULMPC, the CLMPC incorporates torque and momentum constraints.  
Its OCP is formulated as
\begin{equation}\label{eq23}
\begin{aligned}
\operatorname*{argmin}_{\boldsymbol{\mathcal{T}}_c(k)} \quad & \boldsymbol{\mathcal{Y}}^T(k)\boldsymbol{\mathcal{Y}}(k)+W_{CLMPC}\boldsymbol{\mathcal{T}}_c^T(k)\boldsymbol{\mathcal{T}}_c(k) \\
\text{s.t.} \quad & \boldsymbol{T}_{min} \leqslant \boldsymbol{T}_c(k+i) \leqslant \boldsymbol{T}_{max},\\
&i=0, 1, \cdots, N_c-1\\
                 & \boldsymbol{h}_{min} \leqslant \boldsymbol{h}_{rw}(k)-t_s\sum_{j=0}^{i}\boldsymbol{T}_c(k+j) \leqslant \boldsymbol{h}_{max},\\
&i=0, 1, \cdots, N_c-1
\end{aligned}
\end{equation}

\subsection{NMPC}
Unlike the LMPC formulation, the NMPC considers the full nonlinear model and is formulated as the following OCP:
\begin{equation}\label{eq24}
\begin{aligned}
\operatorname*{argmin}_{\boldsymbol{\mathcal{T}}} \quad & \sum_{k=0}^{N-1}\left( \vec{\boldsymbol{q}}_{e,k}^T \vec{\boldsymbol{q}}_{e,k} + W_{NMPC} \boldsymbol{T}^T_k \boldsymbol{T}_k \right) \\
\text{s.t.} \quad & \vec{\boldsymbol{q}}_{e,k+1}=\boldsymbol{f}(\boldsymbol{q}_{e,k},\boldsymbol{\omega}_k,\boldsymbol{T}_k)\\
&\boldsymbol{T}_{min} \leqslant \boldsymbol{T}_k \leqslant \boldsymbol{T}_{max}, k=0, 1, \cdots, N-1\\
& \boldsymbol{h}_{min} \leqslant \boldsymbol{h}_{rw,k} \leqslant \boldsymbol{h}_{max}, k=0, 1, \cdots, N-1
\end{aligned}
\end{equation}
where $*_k$ denotes the variable at time $kt_s$, $\vec{\boldsymbol{q}}_{e}$ is the vector part of the quaternion error $\boldsymbol{q}_e$, $\boldsymbol{\mathcal{T}}=\{\boldsymbol{T}_0,\boldsymbol{T}_1,\ldots,\boldsymbol{T}_{N-1}\}$ is the control sequence, and $\boldsymbol{f}(*)$ represents the discretized form of Eqs~(\ref{eq1}--\ref{eq4}).  
Here, $N$ is the prediction horizon of NMPC.  
The nonlinear OCP is solved using an SQP approach implemented via the ACADO toolkit \cite{ACADO}, with qpOASES employed as the underlying quadratic programming solver \cite{qpOASES}.

\section{Results and Discussion}\label{cpt4}
\subsection{Simulation Scenario}
The simulations were conducted using a dedicated simulator for attitude determination and control system developed by the National University of Singapore (NUS) and can be found in \cite{ULMPC}.
Ten minutes of position and velocity data were obtained from the NUS satellite. Based on these data, together with the orbital parameters provided in \cite{ULMPC}, the time-varying rotation quaternion from the ECI frame to the LVLH frame, denoted as $\boldsymbol{q}_{\rm LVLH/ECI}$, is generated.
To compare the agility of different MPC methods, we define the rotation quaternion $\boldsymbol{q}_{\rm target/LVLH}$, which represents the transformation from the LVLH frame to the target frame.
The target frame is determined by applying rotations to the LVLH frame according to the prescribed latitude and longitude offsets of the target point relative to the subsatellite point on the Earth's surface.
The target quaternion can be obtained as
\begin{equation}\label{eq25}
\boldsymbol{q}_t=\boldsymbol{q}_{\rm target/LVLH}\otimes\boldsymbol{q}_{\rm LVLH/ECI}
\end{equation}

To enable a comprehensive assessment of both agility and stability among different MPC methods, three phases in ten minutes are designed, as illustrated in Fig.~\ref{fig1}.
\begin{itemize}
    \item \textbf{Phase I (0.5--5.5 min):} The satellite sequentially points to ten different locations, issuing a reorientation command every 30 seconds. This scenario is used to simulate rapid imaging of multiple areas.
    \item \textbf{Phase II (6.0--8.0 min):} The satellite continuously tracks a target whose longitude drifts at a rate of 0.67 deg/sec. This scenario is used to simulate rapid ground scanning.
    \item \textbf{Phase III (8.5--9.5 min):} The rate of change of the target's longitude is further increased to 2 deg/sec. This scenario is used to investigate the agility limits of different MPC methods.
    \item \textbf{Others (0.0--0.5 min, 5.5--6.0 min, 8.0--8.5 min, 9.5--10.0 min):} The target frame is aligned with the LVLH frame, i.e., $\Delta {\rm Latitude}=\Delta {\rm Longitude}=0$.
\end{itemize}

The transition intervals between phases (denoted as "Others") are included to allow the satellite to stabilize and to ensure consistent initial conditions for subsequent maneuvers.
For practical considerations, appropriate levels of noise is added to both the sensor and the actuator.
Specifically, the gyroscope measurements are corrupted by additive zero-mean Gaussian white noise with a standard deviation of 0.1 deg/sec.
In addition, the reaction wheel speeds are perturbed by discretized Gaussian noise with a magnitude of 5 rpm.
The detailed simulation environment is summarized in Table~\ref{table2}.
To ensure a fair comparison, all MPC methods are implemented with the same horizon, and their weighting parameters are tuned to achieve similar energy consumption. The specific control parameters used in the simulations are listed in Table~\ref{table3}.
\begin{table}
\renewcommand{\arraystretch}{1.3}
\caption{\textbf{Simulation Environment}}
\label{table2}
\centering
\begin{tabular}{|c|c|c|}
\hline
\bfseries Parameter & \bfseries Value  & \bfseries Unit\\
\hline\hline
$noise_{gyro}$ & 0.27 & $\text{deg}/\text{s}$\\
$noise_{rw}$ & 5 & $\text{rpm}$\\
$t_s$ & 2 & $\text{sec}$\\
$T_{max}$ & $4.18\times10^{-3}$ & $\text{Nm}$\\
$T_{min}$ & $-4.18\times10^{-3}$ & $\text{Nm}$\\
$h_{max}$ & $1.84\times10^{-2}$ & $\text{Nms}$\\
$h_{min}$ & $-1.84\times10^{-2}$ & $\text{Nms}$\\
\hline
\end{tabular}
\end{table}
\begin{table}
\renewcommand{\arraystretch}{1.3}
\caption{\textbf{Controller Parameters}}
\label{table3}
\centering
\begin{tabular}{|c|c|}
\hline
\bfseries Parameter & \bfseries Value\\
\hline\hline
$N_p$ & 10\\
$N_c$ & 10\\
$N$ & 10\\
$W_{A-CLMPC}$ & 40\\
$W_{ULMPC}$ & 10\\
$W_{CLMPC}$ & 10\\
$W_{NMPC}$ & 10\\
\hline
\end{tabular}
\end{table}

\begin{figure*}[!htb]
\centering
\includegraphics[width=\textwidth]{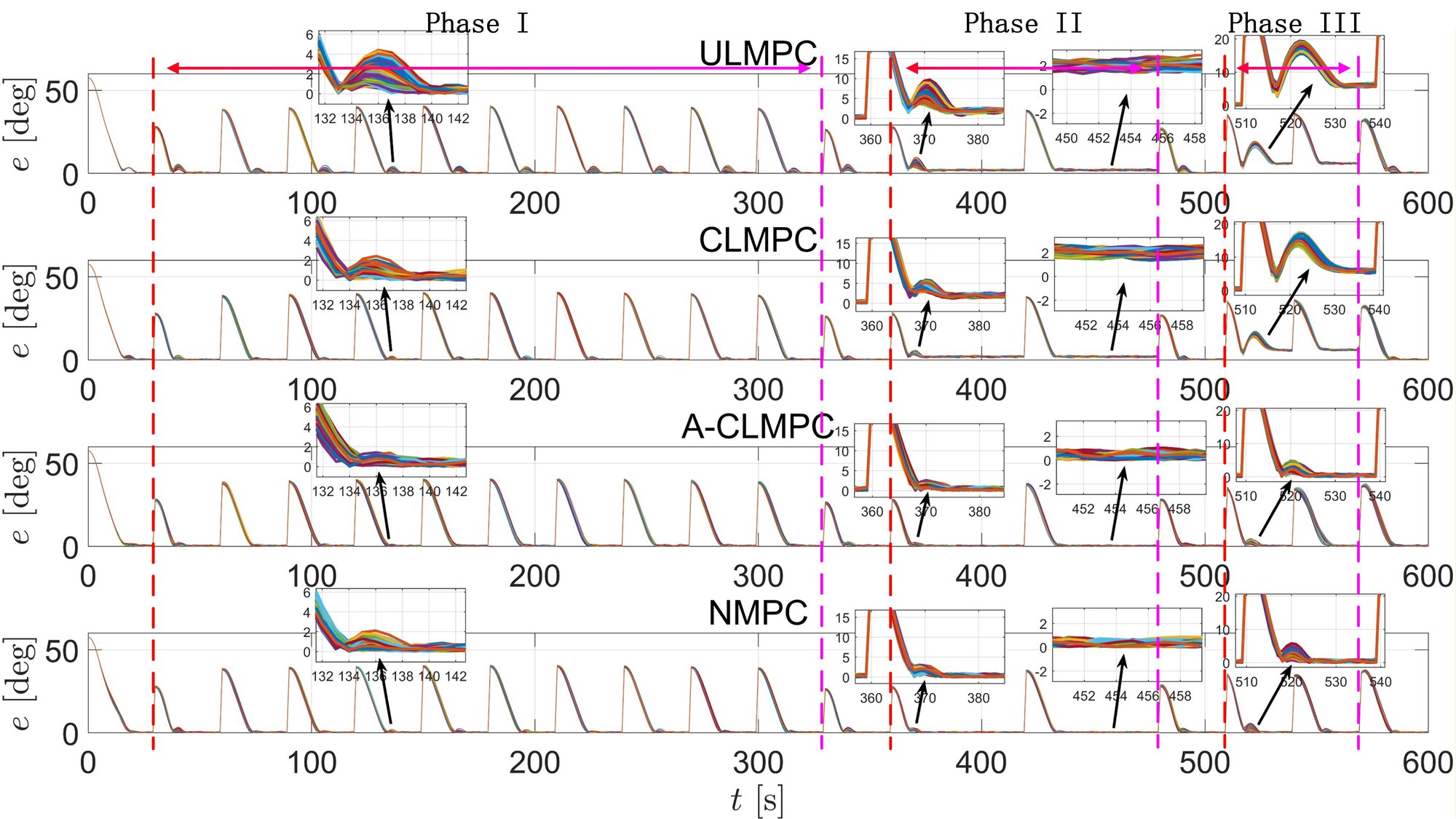}\\
\caption{Attitude error angle derived from quaternion error over 100 Monte Carlo simulations.}
\label{fig2}
\end{figure*}
\begin{table*}[!htb]
\centering
\caption{\textbf{Quantitative results from 100 Monte Carlo simulations.}}
\label{table4}
{\begin{tabular}{lccccccccccccccc}
\hline \midrule
 \multirow[t]{2}{*}{Controller} & \multicolumn{3}{c}{ULMPC} & & \multicolumn{3}{c}{CLMPC} & & \multicolumn{3}{c}{A-CLMPC} & & \multicolumn{3}{c}{NMPC}\\
\cmidrule{2-4} \cmidrule{6-8} \cmidrule{10-12} \cmidrule{14-16}
Phase & I & II & III & & I & II & III & & I & II & III & & I & II & III\\\midrule
TT\textsuperscript{a} [s]
& 16.94 & 12.66 & 14.74 & & 14.16 & 11.79 & 14.19 & & 14.18 & 10.40 & 11.14 & & 13.64 & 10.37 & 10.71
\\\midrule
SSE\textsuperscript{b} [deg]
& 0.33 & 1.80 & 5.43 & & 0.34 & 1.79 & 5.50 & & 0.40 & 0.49 & 1.00 & & 0.33 & 0.50 & 1.16
\\\midrule
MSE\textsuperscript{c} [${\rm deg}^2$]
& 0.14 & 3.33 & 30.00 & & 0.15 & 3.31 & 30.85 & & 0.20 & 0.34 & 1.34 & & 0.15 & 0.32 & 1.73 \\
 \hline \hline
\end{tabular}}\\
\textsuperscript{a}TT= Average transition time per area;\quad
\textsuperscript{b}SSE= Average steady-state error;\quad
\textsuperscript{c}MSE= Mean squared error in the steady-state period.
\end{table*}

\begin{figure*}[!htb]
\vspace{3mm}
\centering
\includegraphics[width=\textwidth]{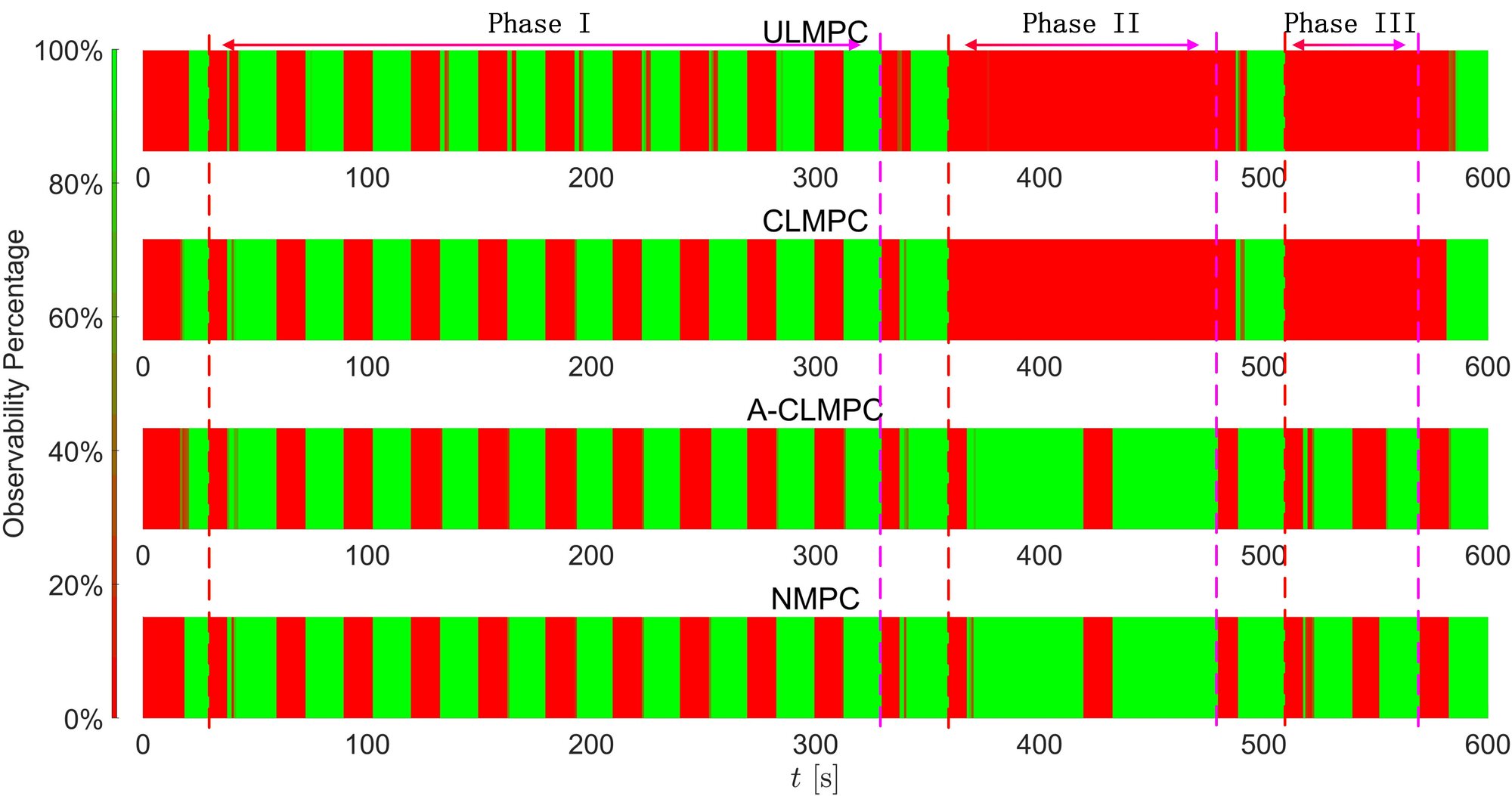}\\
\caption{Target observability percentage over 100 Monte Carlo simulations.}
\label{fig3}
\end{figure*}

\subsection{Simulation Results and Discussion}
Fig.~\ref{fig2} illustrates the attitude tracking error obtained from 100 Monte Carlo simulations, while the corresponding quantitative results are summarized in Table~\ref{table4}.
Throughout all phases, the ULMPC method, which does not account for actuator saturation, exhibits noticeable overshoots during rapid maneuvers and shows the longest transition time among all MPC methods.
When actuator saturation is explicitly considered, CLMPC effectively limits excessive control effort.
However, similar to ULMPC, it still exhibits steady-state errors in Phases II and III, when the target is in motion.
The steady-state error arises from model mismatches in LMPC caused by system linearization and feedforward control, and its magnitude increases with the target's motion speed.
In contrast, the A-CLMPC employs an embedded integrator to compensate for model mismatches. Therefore, the tracking error in Phase II is reduced by 72.22\% compared with ULMPC and by 72.07\% compared with CLMPC, while in Phase III it is reduced by 81.58\% compared with ULMPC and by 81.82\% compared with CLMPC.
Moreover, the A-CLMPC also shortens the transition time compared to other LMPC methods across all phases.
Unlike LMPC-based methods, the NMPC captures the full nonlinear dynamics, achieving a faster transient response and a steady-state error comparable to, or even slightly larger than, that of the proposed A-CLMPC.
Overall, ULMPC is limited by actuator constraints, while CLMPC enhances dynamic performance but still exhibits steady-state errors, whereas A-CLMPC and NMPC demonstrate superior accuracy and agility across all phases.

This paper considers a commercial camera with a swath angle of $0.8022$ deg.
The target is deemed observable when the satellite's attitude tracking error remains within this angular range.
Fig.~\ref{fig3} shows the observability percentages of different MPC methods over 100 Monte Carlo simulations.
The ULMPC exhibits a longer transition time, which reduces the observable duration (with an average of 14.7 sec of full observability per area) during Phase I.
Although CLMPC shortens the transition time (achieving an average of 16.1 sec of full observability per area), it still fails to maintain observability in Phases II and III due to steady-state errors when the target is in motion.
In contrast, the A-CLMPC (achieving an average of 15.7 sec of full observability per area) maintains a high level of observability comparable to NMPC (achieving an average of 16.1 sec of full observability per area).
Both methods achieve nearly continuous target observability across all phases, demonstrating their effectiveness in handling agile tracking tasks.

\begin{figure*}[!htb]
\centering
\begin{subfigure}[b]{0.46\textwidth}
    \includegraphics[width=\linewidth]{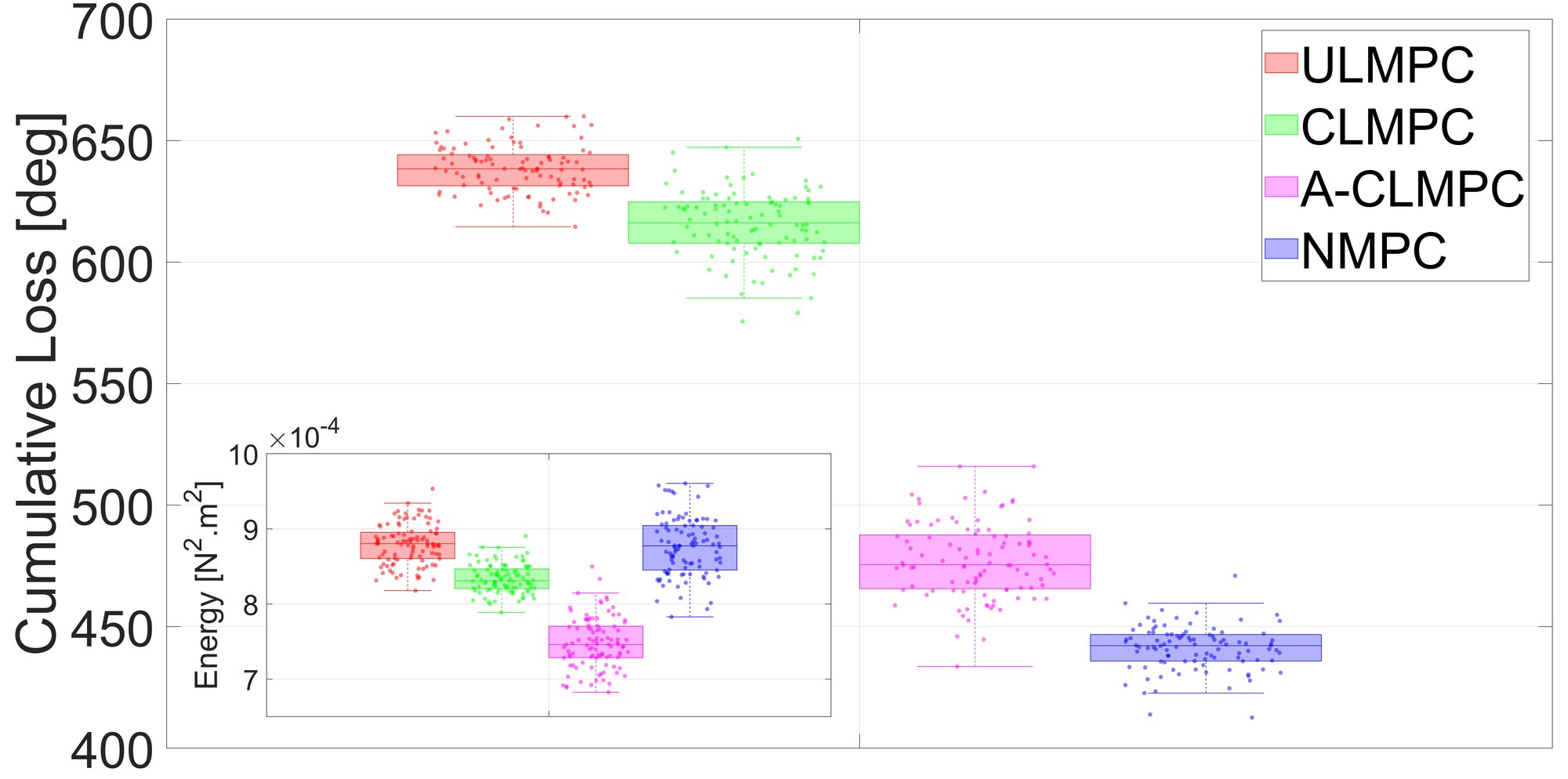}
    \caption{Transient period.}
    \label{fig4-1}
\end{subfigure}
\hspace{0.05\textwidth}
\begin{subfigure}[b]{0.46\textwidth}
    \includegraphics[width=\linewidth]{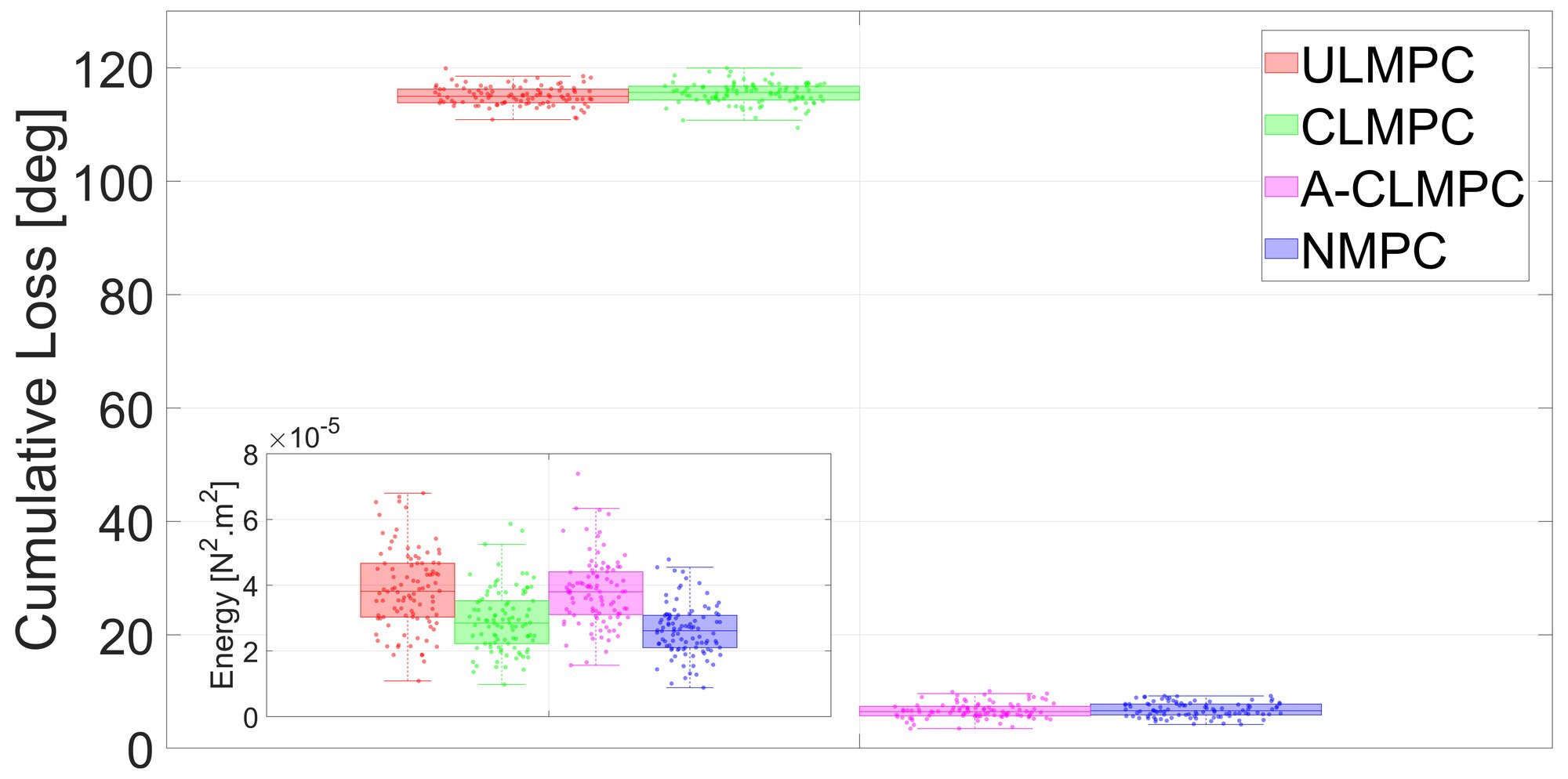}
    \caption{Steady-state period.}
    \label{fig4-2}
\end{subfigure}
\caption{Cumulative loss and energy consumption during phase III.}
\label{fig4}
\end{figure*}

To further assess the reliability of different MPC methods in managing agile maneuvers, Fig.~\ref{fig4} presents the cumulative tracking error and actuator energy consumption during the transition and steady-state periods of Phase III.
ULMPC and CLMPC accumulate more errors in both periods despite using similar amounts of energy (within the neighbourhood of  $8\times10^{-4}\,{\rm N}^2\!\cdot{\rm m}^2$ during the transition period and $3\times10^{-5}\,{\rm N}^2\!\cdot{\rm m}^2$ during the steady-state period), reflecting their limited agility.
The A-CLMPC accumulates 74.48\% and 77.17\% of the transition-period error of ULMPC and CLMPC, respectively, which is only 7.53\% higher than that of NMPC.
In the steady-state period, its cumulative error reduces to 5.61\% and 5.57\% of that of ULMPC and CLMPC, respectively, and matches the performance of NMPC.

In summary, the proposed A-CLMPC demonstrates agility comparable to NMPC while outperforming both ULMPC and CLMPC in both tracking precision and transition performance.
Nevertheless, numerical simulations do not adequately reflect the computational burden of different MPC methods.
Therefore, the following subsection presents experimental results to further evaluate the real-time feasibility of four controllers.

\subsection{Experimental Results and Discussion}
\begin{figure}
\centering
\includegraphics[width=\linewidth]{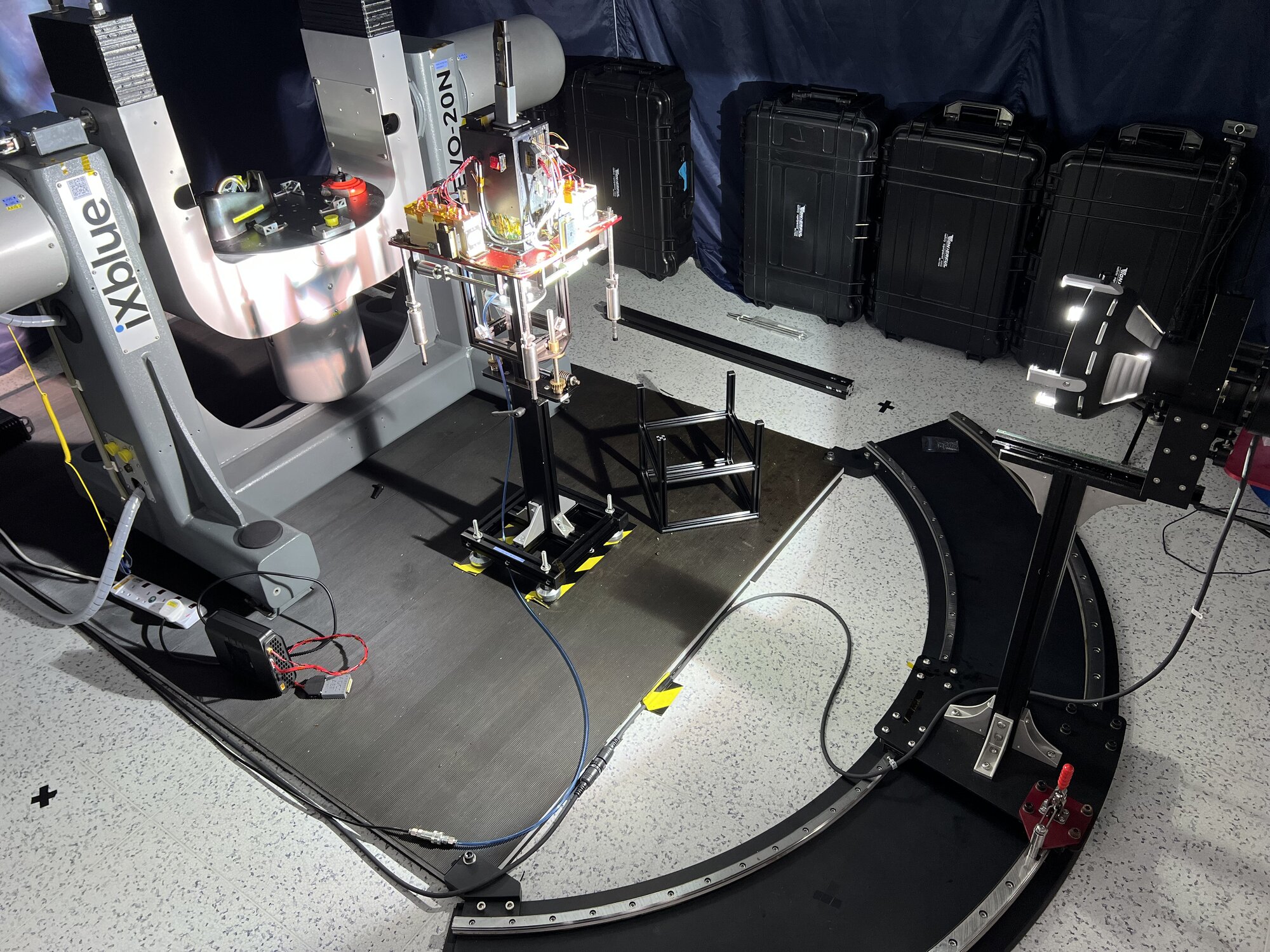}\\
\caption{Experimental platform for satellite attitude determination and control system \cite{ULMPC,future1}.}
\label{fig5}
\end{figure}

The experimental validation was conducted on a satellite attitude control platform mounted on an air-bearing table, which enabled near-frictionless rotational motion, as shown in Fig.~\ref{fig5}.
A light source was used to emulate the Sun, and the satellite was equipped with sun sensors to continuously measure the incident direction of the light.
The yaw reference command is therefore generated according to the relative position of the light source, allowing the satellite to perform continuous yaw tracking.
This setup enables a realistic comparison of different MPC methods in terms of their tracking performance and computational feasibility.

\begin{figure}
\centering
\includegraphics[width=\linewidth]{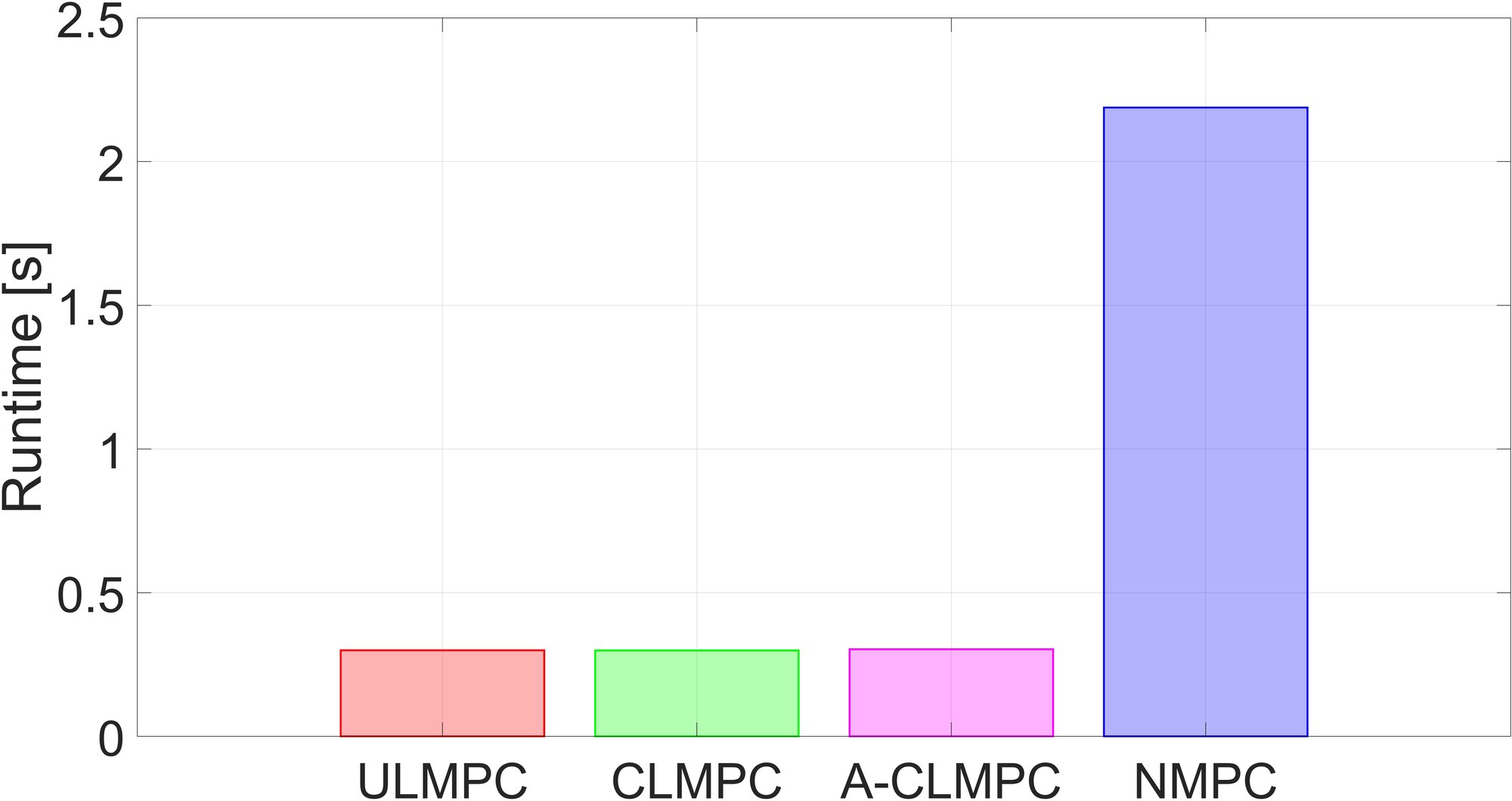}\\
\caption{Runtime required to deploy different MPC methods.}
\label{fig6}
\end{figure}

The NMPC solves OCPs by iteratively performing gradient descent of the cost function.
As a result, considering the full nonlinear dynamics, NMPC requires substantially higher computational resources compared with LMPC, which relies on linearized and simplified models.
Fig.~\ref{fig6} presents the average runtime of each MPC method per control update on the experimental platform.
It can be observed that the A-CLMPC, similar to other LMPC methods, computes the optimal torque within approximately 0.3 sec, whereas NMPC requires about 2.2 sec—exceeding the control sampling period (2 sec).
Such a high computational demand introduces intrinsic delay in generating the optimal torque, preventing NMPC from stably tracking the agile command in real-time experiments, and therefore benchmarking using NMPC was excluded from the experiment.

\begin{figure}
\centering
\includegraphics[width=\linewidth]{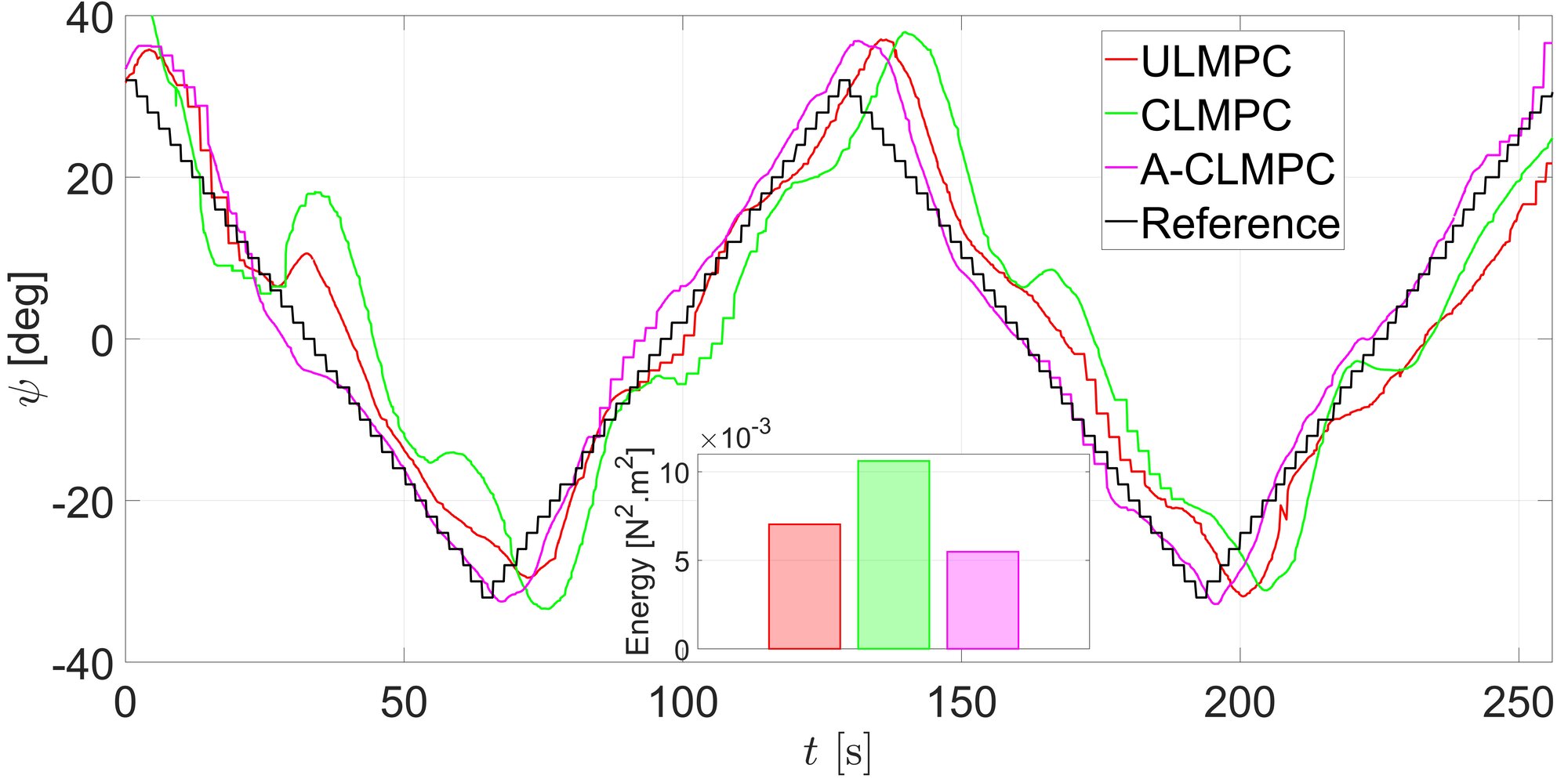}\\
\caption{Yaw tracking performance and energy consumption of different MPC methods.}
\label{fig7}
\end{figure}

In the experiment, the yaw reference command oscillates between $\pm$32 deg at an angular rate of 1 deg/sec.
The tracking performance of different MPC methods is illustrated in Fig.~\ref{fig7}.
As observed, consistent with the numerical simulation results, the proposed A-CLMPC successfully eliminates the steady-state error. Quantitatively, its cumulative tracking error is reduced to 69.9\% of that of ULMPC and 45.4\% of that of CLMPC, while the reaction wheel energy consumption decreases to 77.9\% and 51.7\% of theirs, respectively.


\section{Conclusion}\label{cpt5}
This paper proposes an A-CLMPC algorithm that enables AEOSs to achieve NMPC-level agility without incurring its substantial computational burden.
Comprehensive numerical simulations and physical experiments demonstrate that the proposed controller not only eliminates the steady-state error inherent to conventional LMPCs but also achieves faster transient responses and reduced energy consumption.
These results indicate that the A-CLMPC provides a practical balance between agility and computational complexity.
The proposed method establishes an upper bound on the achievable agility of small satellites, which are often constrained by limited size, weight, and power, and therefore lack the computational resources required by NMPC.
It enables effective agile tracking even with limited actuation capability.
Future work will focus on extending this framework to on-orbit applications and further exploring the agility potential of resource-constrained satellites \cite{future1}.


\begin{thebibliography}{99}

\bibitem{intro2} G. Soldi, D. Gaglione, N. Forti, A. D. Simone, F. C. Daffin`a, G. Bottini, D. Quattrociocchi, L. M. Millefiori, P. Braca, S. Carniel, P. Willett, A. Iodice, D. Riccio, and A. Farina, "Space-based global maritime surveillance. part i: Satellite technologies," IEEE Aerospace
and Electronic Systems Magazine, vol. 36, no. 9, pp. 8–28, 2021.
\bibitem{intro1} M. N. Sweeting, "Modern small satellites-changing the economics of space," Proceedings of the IEEE, vol. 106, no. 3, pp. 343–361, 2018.
\bibitem{actuator1} L. O. Inumoh, N. M. Horri, J. L. Forshaw, and A. Pechev, "Bounded gain-scheduled lqr satellite control using a tilted wheel," IEEE Transactions on Aerospace and Electronic Systems, vol. 50, no. 3, pp. 1726– 1738, 2014.
\bibitem{scheduling1} X. Wang, G. Wu, L. Xing, and W. Pedrycz, "Agile earth observation satellite scheduling over 20 years: Formulations, methods, and future directions," IEEE Systems Journal, vol. 15, no. 3, pp. 3881–3892, 2021.
\bibitem{scheduling2} W. Lu, W. Gao, B. Liu, W. Niu, D. Wang, Y. Li, X. Peng, and Z. Yang, "Reinforcement learning driven time-sensitive moving target tracking of intelligent agile satellite," IEEE Transactions on Aerospace and Electronic Systems, vol. 60, no. 6, pp. 9085–9101, 2024.
\bibitem{LMPC1} L. Zhao, Z. Lu, K. V. Ling, Y. Hu, K. Zheng, and W. Liao, "Multi-rate cascade spacecraft attitude-orbit integrated state estimation and control framework based on mhe and pwa-mpc," IEEE Transactions on Aerospace and Electronic Systems, pp. 1–19, 2025.
\bibitem{LMPC2} R. J. Caverly, S. D. Cairano, and A. Weiss, "Electric satellite station keeping, attitude control, and momentum management by mpc," IEEE Transactions on Control Systems Technology, vol. 29, no. 4, pp. 1475– 1489, 2021.
\bibitem{add20250909} M. AlandiHallaj and N. Assadian, "Multiple-horizon multiple-model predictive control of electromagnetic tethered satellite system," Acta Astronautica, vol. 157, pp. 250–262, 2019.
\bibitem{nmpc1} Y. Zhou, Y. Hu, K.V. Ling, and F. Ding, "Hybrid two-stage identification-based nonlinear mpc strategy for satellite attitude control," IEEE Transactions on Aerospace and Electronic Systems, pp. 1–12, 2025.
\bibitem{RLvsOC} Y. Song, A. Romero, M. Muller, V. Koltun, and D. Scaramuzza, "Reaching the limit in autonomous racing: Optimal control versus reinforcement learning," Science Robotics, vol. 8, no. 82, p. eadg1462, 2023.
\bibitem{LMPCvsNMPC} M. Kamel, M. Burri, and R. Siegwart, "Linear vs nonlinear mpc for trajectory tracking applied to rotary wing micro aerial vehicles," IFAC-PapersOnLine, vol. 50, no. 1, pp. 3463–3469, 2017, 20th IFAC World Congress.
\bibitem{ECC1} D. Xu and M. Lazar, "Fast model predictive control of power amplifiers for nanometer precision motion systems," 2023 European Control Conference (ECC), Bucharest, Romania, 2023, pp. 1-7.
\bibitem{ECC2} P. Krupa, J. Camara, I. Alvarado, D. Limon and T. Alamo, "Real-time implementation of MPC for tracking in embedded systems: Application to a two-wheeled inverted pendulum," 2021 European Control Conference (ECC), Delft, Netherlands, 2021, pp. 669-674.
\bibitem{book1} B. Stevens, F. Lewis, and E. Johnson, Aircraft Control and Simulation. Wiley, 2016.
\bibitem{ULMPC} M. S. C. Tissera, K. J. E. Foo, K.S. Low, S. T. Goh, and R. D. Tan, "Roekf-mpc estimator for satellite attitude and gyroscope bias estimation," IEEE Transactions on Aerospace and Electronic Systems, vol. 59, no. 5, pp. 4870–4882, 2023.
\bibitem{book2} L. Wang, Model Predictive Control System Design and Implementation Using MATLAB. Springer, 2009.
\bibitem{qpOASES} H. J. Ferreau, C. Kirches, A. Potschka, H. G. Bock, and M. Diehl, "qpoases: A parametric active-set algorithm for quadratic programming," Mathematical Programming Computation, vol. 6, pp. 327–363, 2014.
\bibitem{ACADO} R. Verschueren, G. Frison, D. Kouzoupis, N. van Duijkeren, A. Zanelli, R. Quirynen, and M. Diehl, "Towards a modular software package for embedded optimization," IFAC-PapersOnLine, vol. 51,
no. 20, pp. 374–380, 2018, 6th IFAC Conference on Nonlinear Model Predictive Control NMPC 2018.
\bibitem{future1} K. J. E. Foo, M. S. C. Tissera, K. S. Low, and A. Srivastava, "Flitesim: A comprehensive verification and validation environment for small satellite attitude determination and control systems," IEEE Access, vol. 13, pp. 109 069–109 086, 2025.
\end{thebibliography}
\end{document}